\documentclass{article}
\usepackage{spconf,amsmath,graphicx}
\usepackage{multirow}
\usepackage[noend]{algpseudocode}
\usepackage{algorithm}
\usepackage{cite}
\usepackage{amssymb,amsfonts}
\usepackage[caption=false,font=footnotesize]{subfig}

\title{Service Provisioning and Profit Maximization in Network-assisted Adaptive HTTP Streaming}
\name{Zhisheng Yan$^1$, Cedric Westphal$^{2,3}$, Xin Wang$^2$, and Chang Wen Chen$^1$} 
\address{$^1$State University of New York at Buffalo, $^2$Huawei US R\&D Center, $^3$UCSC
        }

\begin{document}
%
\maketitle
\begin{abstract}
\vspace{-1ex}
Adaptive HTTP streaming with centralized consideration of multiple streams has gained increasing interest. It poses a special challenge that the interests of both content provider and network operator need to be deliberately balanced. More importantly, the adaptation strategy is required to be flexible enough to be ported to various systems that work under different network environments, QoE levels, and economic objectives. To address these challenges, we propose a Markov Decision Process (MDP) based network-assisted adaptation framework, wherein cost of buffering, significant playback variation, bandwidth management and income of playback are jointly investigated. We then demonstrate its promising service provisioning and maximal profit for a mobile network in which fair or differentiated service is required.
\vspace{-1.5ex}
\end{abstract}
%
%
\section{Introduction}\label{sec1}
\vspace{-1.5ex}
Thanks to strong scalability and versatility, HTTP adaptive streaming has established itself as the dominant technique for Internet video delivery, and is expected to stay as a major component of video delivery in the future Internet. 
The most challenging research task for HTTP adaptive streaming is the rate selection process, that is, which quality version of a segment should be streamed. Many works \cite{PANDA14,MDPGLOBE12} have been proposed towards the optimal adaptation conducted by the client itself using local measurements and estimation. Although these rate adaptation algorithms showed promising performance in single-user networks, they cannot be directly applied to multi-client mobile networks with shared bandwidth bottleneck. Playback instability and unfairness have been commonly identified as among the weaknesses for client-side adaptation in practical experiments where two or more clients compete for the same dynamic bottleneck \cite{baseline12,servershape13}. In fact, the fundamental problem behind these issues is that client-side adaptation naively assumes the wireless network is utilized by the client itself and has no knowledge of other competing streams in the shared bottleneck. 

Therefore, it is natural for the network operator to arbitrate such competition and globally adapt the clients' decisions by considering multiple players' status and the overall network environment. MPEG and 3GPP have included such ideas into its working draft \cite{MPEGcore14} or technical report \cite{3GPP938} in order to further enhance the quality of experience (QoE) of Internet streaming. In addition to performance improvement, there are also economic incentives for a network operator to join the adaptive streaming ecosystem.
For instance, content provider \emph{Netflix} has recently entered a peering agreement on smooth streaming with network operator \emph{Comcast} \cite{netflixpay}. 
One of the most critical challenge, however, is how to balance the interests of a content provider and a network operator. Indeed, a content provider always wants its users to experience satisfactory QoE whereas a mobile operator sometimes might be more concerned about its bandwidth cost under congestion. Hence, the adaptation strategy shall be able to guarantee the users' QoE, as well as maximizing the operator's profit. On top of this important challenge, it is also crucial to design the joint adaptation framework in a flexible and systematical way such that the framework can be easily ported to various applications and network environments. For example, there are diverse types of service provision, such as fair experience versus differentiated QoE, high-quality playback versus highly-smooth playback, and different cost models of network operators, such as bandwidth-centric versus user-number-centric.

Only a few research works have been focused on Internet streaming with competing clients. Some exploited traffic shaping mechanisms in the server side to orchestrate the experience of two competing users \cite{baseline12,servershape13}. Others aimed at optimizing various kinds of QoE utility, such as rate-resolution utility \cite{attsand13} and QoE continuum \cite{coe14}, by developing individual optimal or heuristic algorithm. One unique approach called WiDASH \cite{widash12} inserts an additional proxy in between the server and the wireless access networks to implement a split-TCP architecture. 
Nonetheless, all these works concentrate on a specific QoE objective only from the perspective of content providers, which limits their applicability. 

The major contribution of this research is in the development of a \emph{network-assisted} adaptive HTTP streaming framework for multiple clients competing for shared bandwidth over the same bottleneck link. The scheme can flexibly achieve different system-level service requirements by tuning certain framework parameters and can maximize the economic profit of the operator at the same time. Specifically, we propose a Markov Decision Process (MDP) based framework to carry out the adaptation decision, where we jointly consider the clients' bandwidth and service requirements, and thoroughly investigates the operator's economic benefits. 
In the following, we focus on mobile cellular networks to introduce the framework and we describe the system from the point of view of mobile operators. Then we demonstrate two case studies where service fairness and service differentiation (for premium users and regular users) is required by a content provider within an operator-controlled mobile network. These have been identified as major use cases of network-assisted streaming by MPEG \cite{MPEGcore14}. However, it should be noted that the applicability of the framework is general and can be properly adapted to other service requirements.

\begin{figure}[!t]
\centering
\includegraphics[width=3.2in]{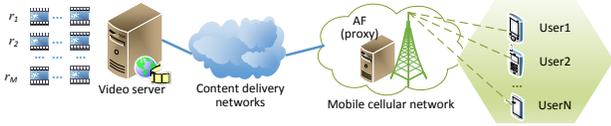}
\vspace{-2ex}
\caption{Architecture of the proposed mobile streaming system.}
\label{sysarch}
\vspace{-2ex}
\end{figure}

\vspace{-2ex}
\section{MDP-based Network-assisted Adaptation Framework}\label{sec4}
\vspace{-1.25ex}
We consider a cellular Internet streaming system as shown in Fig. \ref{sysarch}. We target one cell, where $\mathcal{U}$
is the set of users and each user is indexed by $i = 1, 2, \cdots, N$. There are a set of videos $\mathcal{V}$ with different quality, characterized by bit-rate $r_j$ where $j=1, 2, \cdots, M$. Each quality version is split into $T_{seg}$-second segments. We shift the intelligence from clients to a logically centralized controller within mobile operator's network, i.e., the Application Function (AF) within 3GPP networks \cite{3GPP938}, for the network-assisted adaptation. 

The proposed system works as follows. The video server initially sends out a description of video versions. At each switching point that occurs every $T_{seg}$ seconds, a user requests a segment based on local throughput measurements that need very low complexity. Such request only serves as a \emph{preliminary suggestion} and therefore no corresponding local optimization in the client is carried out. In other words, there would be no conflict if this request is modified later. Unlike conventional client-side adaptation where the mobile operator simply forwards the client requests to the video server, the mobile operator here 
will re-write the request based on MDP-based cell-wide adaptation done by the AF. The client feedback for adaptation, such as available video representations and bandwidth measurement, can be feasibly communicated to the AF. For instance, it can be achieved by standardized quality metrics reporting process \cite{dashstandard} or new techniques like URL parameter insertion\cite{url14}. The final adaptation decision then will be delivered to the cellular scheduler and the video server. We adopt a rate proportional scheduling wherein the downlink resources are allocated based on the upper layer video rate. That way, the video version decided from system's perspective can be effectively streamed to the users.

\vspace{-1.75ex}
\subsection{Wireless Bandwidth Model}
\vspace{-0.75ex}
We model the last-hop cellular link for each client as a Markov process, 
which has been widely studied and proved to be effective \cite{FSMC95}. Each link has $K$ states of available bandwidth, namely $\mathcal{B}=\{bw_k|k=1, 2, \cdots, K\}$. 

The clients measure their own downlink bandwidth and periodically report it to AF, which serves as the input to the adaptation framework. Note that the measured bandwidth may not be exactly the value of $bw_k$. We thereby divide the available bandwidth domain into $K$ regions and map the real bandwidth to the corresponding region. 

\vspace{-1.5ex}
\subsection{Mobile Operator Profit Model}
\vspace{-0.75ex}

\emph{\textbf{Normal Playback Income}.}
At switching point $t$, the mobile operator is expected to obtain income $I_{play}$ from the content provider if it can guarantee a normal playback with rate $R_{t+1}$ during the upcoming period. Formally, this implies $R_{t+1} \leq BW_{t+1}$, where $R_{t+1} \in \mathcal{V}$ and $BW_{t+1} \in \mathcal{B}$ is the estimated bandwidth of link. Considering the logarithmic relationship in rate distortion theory, we model the income from client $i$ as
\begin{equation}\label{i_play}
\vspace{-0.25ex}
I_{i,play}=\Gamma_{play} \alpha \frac{\log{\frac{R_{i,t+1}}{R_{min}}}}{\eta_{play}}
\vspace{-0.25ex}
\end{equation}
where $\alpha$ ($0 \leq \alpha \leq 1$) is the sharing weight that normal playback places on QoE and $R_{min}=\min\{r_j|r_j \in \mathcal{V}\}$. We denote the normalization factor as $\eta_{play}=\log{\frac{R_{max}}{R_{min}}}$, where $R_{max}=\max\{r_j|r_j \in \mathcal{V}\}$. $\Gamma_{play}$ is an indicator function for the normal playback, namely $\Gamma_{play}= 1$ if $R_{t+1} \leq BW_{t+1}$ and $\Gamma_{play}= 0$ otherwise.

\emph{\textbf{Playback Buffering Cost.}}
If the operator-adapted bit-rate exceeds the available bandwidth, playback buffering would occur, which should not be encouraged in the service contract. Hence, we impose certain penalty cost $C_{buf}$ on player $i$.
\begin{equation}\label{c_buf}
\vspace{-0.05ex}
C_{i,buf}=(1-\Gamma_{play})\beta \frac{\log{\frac{R_{i,t+1}-BW_{i,t+1}}{\delta_{min}}}}{\eta_{buf}}
\vspace{-0.05ex}
\end{equation}
where $\beta$ ($0 \leq \beta \leq 1$) is the re-buffering weight, 
and $\delta_{min}=\min\{r_j-bw_k|r_j \in \mathcal{V}, bw_k \in \mathcal{B}\}$. The normalization factor of playback buffering is $\eta_{buf}=\log{\frac{R_{max}-BW_{min}}{\delta_{min}}}$, where $BW_{min}=\min\{bw_k|bw_k \in \mathcal{B}\}$.

\emph{\textbf{Playback Smoothness.}}
If the rate variation in two consecutive periods is too significant, the content provider would penalize the mobile operator some money $C_{var}$. 
\begin{equation}\label{c_var}
C_{i,var}=\Gamma_{var}\gamma \frac{\log{\frac{|R_{i,t}-R_{i,t+1}|}{\Delta}}}{\eta_{var}}
\end{equation}
where $\gamma$ ($0 \leq \gamma \leq 1$) is the smoothness weight and $\alpha+\beta+\gamma=1$, $\Delta$ is the rate variation threshold below which no penalty would occur, and $\eta_{var}=\log{\frac{R_{max}-R_{min}}{\Delta}}$ is the normalization factor. We can thus express the indicator function $\Gamma_{var}$ as, $\Gamma_{var}= 1$ if $R_{t}-R_{t+1} \geq \Delta$ and $\Gamma_{var}= 0$ otherwise.

\textbf{\emph{Bottleneck Bandwidth Cost.}}
Finally, 
the mobile cellular network may experience radio congestion, during which the mobile operator would have to spend proportionally more bandwidth cost to manage the network: 
\begin{equation}\label{c_bw}
\vspace{-0.25ex}
C_{bw}=\Gamma_{bw}\theta(\sum_{i\in\mathcal{U}} R_{i,t+1}-R_{th})
\end{equation}
where $R_{th}$ is the total service rate constraint above which the mobile operator has to spend extra money, $\theta$ is the cost per unit exceeded rate, $\Gamma_{bw}=1$ when $\sum_{i\in\mathcal{U}} R_{i,t+1}>R_{th}$, and otherwise $\Gamma_{bw}=0$.

\vspace{-1.5ex}
\subsection{Markov Decision Process Formulation}
\vspace{-0.7ex}
MDP is a 4-tuple reinforcement learning task \cite{ReLear} that can intelligently interact with the uncertain bandwidth estimation and make the adaptation decision accordingly.

\textbf{\emph{System States.}}
The system state at time period $t$ can be defined as $s_t=(\mathbf{R_t},\mathbf{BW_t})$, where the vector represents the bandwidth and quality version for all clients in $\mathcal{U}$. Besides, the system state set $\mathcal{S}=\{s_1, s_2, \cdots, s_T\}$ essentially keeps track of the evolution of the entire streaming system, where $T$ is session duration. Since the system state depends only on its most recent (previous) state, the Markov property holds.

\textbf{\emph{System Actions.}}
The action set $\mathcal{A}=\{a_1, a_2, \cdots, a_T\}$ defines the operator-selected quality version of video for all the users. In particular, $a_t=\mathbf{R_{t+1}}$, denotes the quality version to be streamed in the upcoming adaptation period.

\textbf{\emph{State Transitions.}}
The state transition from $s_t$ to $s_{t+1}$ is determined by the decided quality version and available bandwidth at time period $t$. 
Given that the mobility-decided bandwidth variation is independent of bit-rate decisions, the state transition probability $\mathcal{P}_{a_t}(s_t,s_{t+1})$ can be derived as,
\begin{equation}\label{p_tx}
\vspace{-0.15ex}
\begin{split}
\mathcal{P}_{a_t}(s_t,s_{t+1}) &= \text{Pr}\{s_{t+1}|s_t\} \\
                               &=\text{Pr}\{(\mathbf{R_{t+1}},\mathbf{BW_{t+1}})|(\mathbf{R_{t}},\mathbf{BW_{t}})\} \\
                               &=\text{Pr}\{\mathbf{R_{t+1}}|\mathbf{R_t}, \mathbf{R_{t+1}}=a_t\} \text{Pr}\{\mathbf{BW_{t+1}}|\mathbf{BW_{t}}\}
\end{split}
\vspace{-0.3ex}
\end{equation}
where $\text{Pr}\{\mathbf{BW_{t+1}}|\mathbf{BW_{t}}\}$ can be obtained from the transition matrix of the channel model and $\text{Pr}\{\mathbf{R_{t+1}}|\mathbf{R_t}, \mathbf{R_{t+1}}=a_t\}$ is decided by the action policy.

\textbf{\emph{Profit Function and Optimization Objective.}}
The profit function $\mathcal{R}_{a_t}(s_t,s_{t+1})$ is the overall monetary profit of the mobile operator from $s
_t$ to $s_{t+1}$, 
i.e.,
\begin{equation}\label{profit}
\vspace{-0.05ex}
\mathcal{R}_{a_t}(s_t,s_{t+1})=\sum_{i \in \mathcal{U}}\lambda_i(I_{i,play}-C_{i,buf}-C_{i,var})-C_{bw}
\vspace{-0.05ex}
\end{equation}
where $\lambda_i$ is the service priority coefficient for user $i$ and $\sum_{i\in\mathcal{U}}\lambda_i=1$. Hence, the optimization objective of the network-assisted streaming is that we seek to take the optimal policy $\pi$ such that the profit of mobile operator during the $T$-second session can be maximized. Thus the objective is $\mathcal{OBJ}=\max_{\pi} \sum_{t=0}^{t=T}{\mathcal{R}_{a_t}(s_t,s_{t+1})}$.

\textbf{\emph{Framework Flexibility.}}
All the above models can be customized for the mobile operator or content providers. Importantly, these modified frameworks can still be optimally solved as introduced next because they only lead to a different profit function $\mathcal{R}_{a_t}(s_t,s_{t+1})$.

\vspace{-1.45ex}
\subsection{MDP Solution}
\vspace{-0.7ex}
We propose a dynamic programming based algorithm to solve the profit-maximized rate adaptation problem. We first let $v(s_t)$ be the maximal expected profit from $s_t$ to the end state $s_T$. Based on Bellman value iteration \cite{ReLear}, we have
\begin{equation}\label{Value}
\vspace{-0.75ex}
    v(s_t)=\max_{a_t}\{\sum_{s_{t+1}\in \mathbb{S}}\mathcal{P}_{a_t}(s_t,s_{t+1})(\mathcal{R}_{a_t}(s_t,s_{t+1})+ v(s_{t+1}))\}
\vspace{-0.1ex}
\end{equation}
where $\mathbb{S}=\mathcal{V}^N\times\mathcal{B}^N$ is the state space. This essentially means, for a given current state $s_t\in \mathbb{S}$ and a given action $a_t$, we have multiple possible next states $s_{t+1}\in \mathbb{S}$ and need to compute the expected rewards for all the transitions. Then we can obtain the optimal rewards by selecting the action achieving the highest expected rewards. By substituting $v(s_T)=0$, the iteration will have a initial value. Iteratively, we can compute the optimal $v(s_{0})$ (equal to the objective $\mathcal{OBJ}$) for all possible states $s_{0}\in \mathbb{S}$ and accordingly the optimal policy.
\begin{equation}\label{pi}
\vspace{-1.25ex}
    \pi=\text{arg}\max\limits_{a}\{\sum_{s_{1}\in \mathbb{S}}\mathcal{P}_{a}(s_0,s_{1})(\mathcal{R}_{a}(s_0,s_{1})+ v(s_{1}))\}
\end{equation}

\vspace{-2.5ex}
\begin{algorithm}
\caption{Dynamic Programming Adaptation Algorithm}\label{dp1}
\begin{algorithmic}[1]
\State \textbf{Initialization}: $t \gets T-1$, $v(s_{t+1})\gets 0$
\While {$t \geq 0$}
    \For {all possible $s_{t} \in \mathbb{S}$}
        \State $v(s_t)\gets\max_{a_t}\{\sum_{s_{t+1}\in \mathbb{S}}\mathcal{P}_{a_t}(s_t,s_{t+1})(\mathcal{R}_{a_t}(s_t,s_{t+1})$ \\$\qquad \quad +v(s_{t+1}))\}$
        \State $v(s_{t+1}) \gets v(s_t)$
    \EndFor
    \State $t \gets t-1$
\EndWhile
\State Output the optimal deterministic policy $\pi$.
\end{algorithmic}
\vspace{-0.75ex}
\end{algorithm}
\vspace{-1.5ex}

An table that maps each state to the optimal quality version would be generated. Then the mobile operator can make the decision by looking up the table at each switching point. We summarize the iterative algorithm in Algorithm \ref{dp1}.

\vspace{-1.65ex}
\section{Performance Evaluations}\label{sec5}
\vspace{-1.6ex}
We evaluate the proposed systems using two case studies, i.e., fair services and differentiated services, with two competing users using \emph{ns-2}. We use H.264/AVC encoded sequence, ``Big Buck Bunny". 
The bit-rate of 5 quality levels is 95.11, 183.53, 364.63, 493.02, and 798.09 Kbps, respectively. 

We use a four-state Markov last-hop channel. Each state lasts for 1 second. We also select segment length $T_{seg}$ to be 1 second to correspond to the channel coherence time. Since we are interested in the service provisioning by the mobile operator, we adopt the same Markov model for all users to eliminate the impacts of channel differentiation. We adopt a Markov channel matrix used in \cite{MDPGLOBE12} to simulate the cellular link, i.e., $\mathbf{P_t}=[.5, .5, 0, 0; .2, .6, .2, 0; 0, .1, .7, .2; 0, 0, .2, .8]$. The bandwidth region boundaries for the four state are 256, 512, and 896 Kbps. 
We use the core network settings in \cite{widash12}.

Regarding the profit model, we assume the overall service rate constraint $R_{th}$ is 850 Kbps. The congestion penalty $\theta$ is set to $\infty$, which indicates that aggressive bit-rate selection is completely forbidden. We set the sharing weight $\alpha$, $\beta$ and $\gamma$ as 0.3, 0.5, and 0.2, respectively, for higher emphasis of buffering. We have run simulations with different sets of weights, and found them insensitive to service provisioning. However, the actual profit of operator would be varied, which indicates mobile operators' potential tradeoff between profit and service agreement. The threshold for rate consistency $\Delta$ is set to be 350 Kbps. The service priority coefficient $\lambda$ is set to be 0.5/0.5 and 0.7/0.3 for the two users in the fair and differentiated services case, respectively. The initial quality are the lowest version and the player's buffer is set to be 80 frames. The video session runs 200 seconds for 15 times.

\vspace{-1.5ex}
\subsection{Fair Services Provisioning}
\vspace{-0.85ex}
In this section, we evaluate the performance of different adaptation algorithms when applied in the network-assisted system, where fair service provisioning is required:\\
$\textrm{i})$ {\bf Upper bound algorithm} (referred as \emph{Ideal}): Same as Algorithm \ref{dp1}, except that future channel condition is perfectly known by the mobile operator such that the MDP transition probability becomes 1.0 when considering an action. Thus we can compute the actual rather than the expected rewards.\\
$\textrm{ii})$ {\bf Myopic algorithm} (referred as \emph{Myopic}): The network-assisted system directly employs the local throughput-based adaptation decision from clients.

\begin{figure}[!t]
\vspace{-2ex}
\centerline{
\subfloat[] {\includegraphics[width=1.75in]{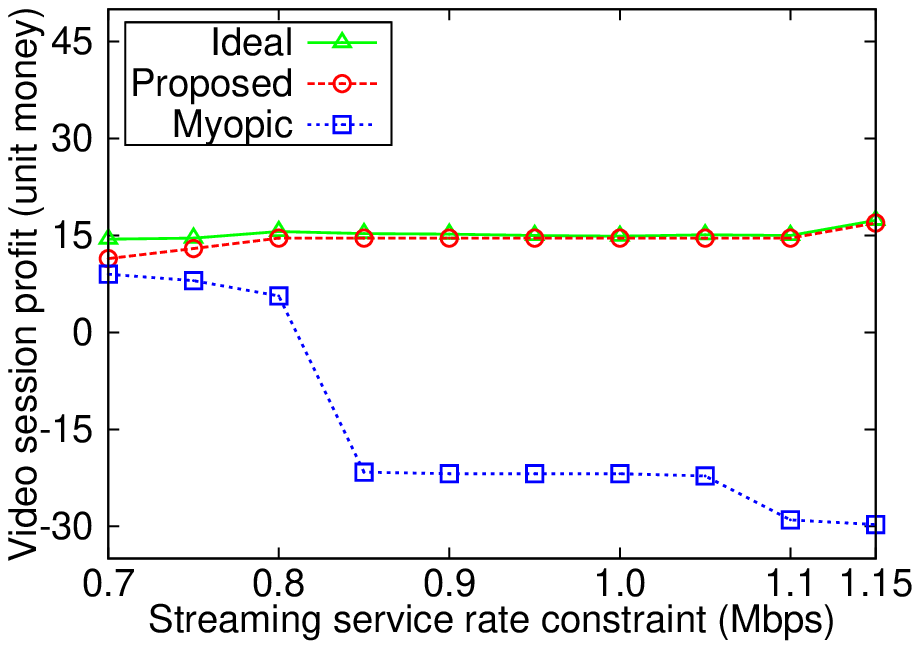}%
\label{pro_gbr}}
\hfil
\subfloat[]{\includegraphics[width=1.75in]{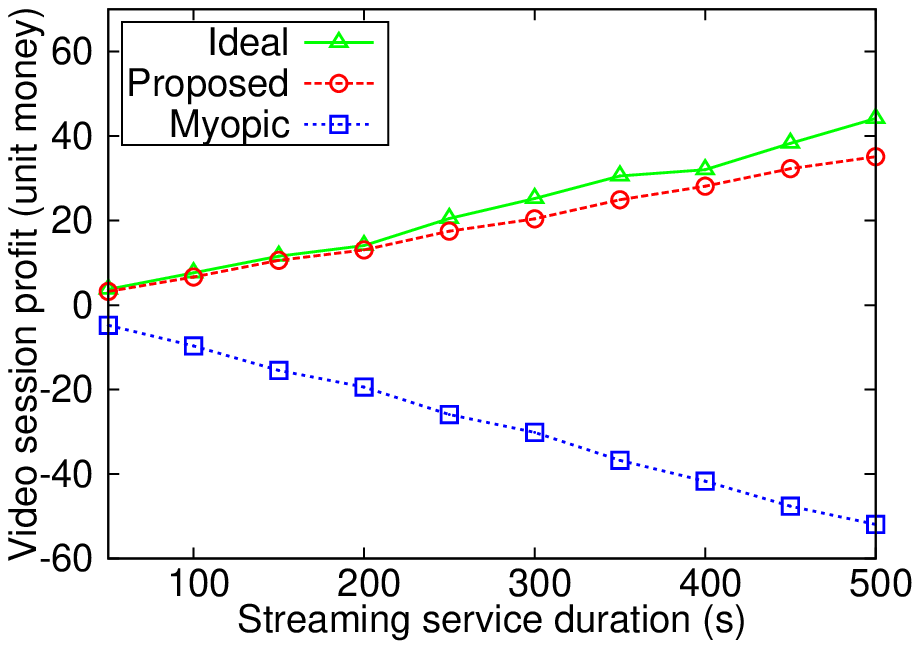}%
\label{pro_t}}}
\vspace{-2.5ex}
\caption{The session profit of mobile operator versus a) service rate constraint $R_{th}$ during congestion; b) session duration $T$.}
\label{pro}
\vspace{-2.85ex}
\end{figure}

We show the total profit of mobile operator versus service rate constraint in Fig. \ref{pro_gbr}. We also demonstrate the profit versus session durations in Fig. \ref{pro_t}. As shown in the figures, the proposed algorithm achieves a similar performance as \emph{Ideal}. This actually demonstrates the insensitivity of the proposed algorithm to channel model accuracy, which generalizes its applicability. Additionally, the proposed algorithm shows a superior performance over \emph{Myopic} because \emph{Myopic}, without studying the profit factors, may decide a high bit-rate that causes client re-buffering or too significant rate variations.

\vspace{-1.6ex}
\subsection{Differentiated Services Provisioning}
\vspace{-0.85ex}
We now evaluate the performance of proposed system under the requirement of service differentiation. 
Since we obtained a similar profit comparison curve as Fig. \ref{pro} for the differentiated services case, we will focus on evaluating other playback metrics, compared with conventional client-centric system.

\begin{figure}[!t]
\vspace{-2ex}
\centerline{
\subfloat[] {\includegraphics[width=1.75in]{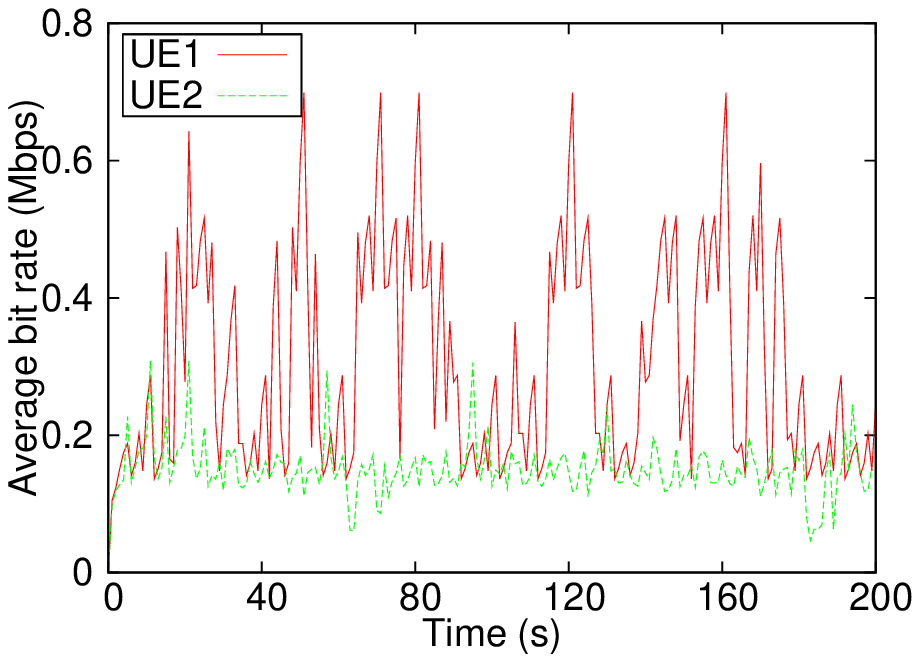}%
\label{abr_p}}
\hfil
\subfloat[]{\includegraphics[width=1.75in]{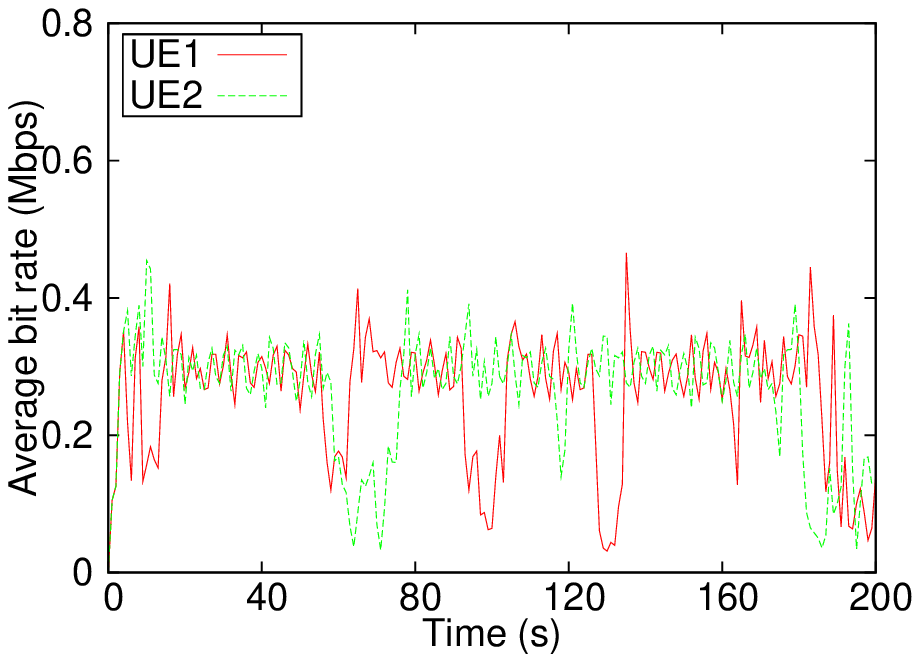}%
\label{abr_c}}}
\vspace{-2.5ex}
\caption{Playback bit-rate versus time of a) the proposed system; b) conventional client-centric system.}
\label{abr}
\vspace{-4ex}
\end{figure}

Fig. \ref{abr} shows playback bit-rate in one session. The high-priority user (UE1) achieves a significant higher rate in the network-assisted system. This is because the mobile operator consider the differentiated service by tuning the priority coefficient, accordingly making service-aware adaptation. However, users enjoy a similar playback rate in client-centric system since they essentially compete for the shared bandwidth. We also found that significant rate variations ($>$350 Kbps) for the proposed system takes more than 5 segments and there is no variations within one segment exceeding the threshold.

\begin{figure}[!t]
\centerline{
\subfloat[] {\includegraphics[width=1.75in]{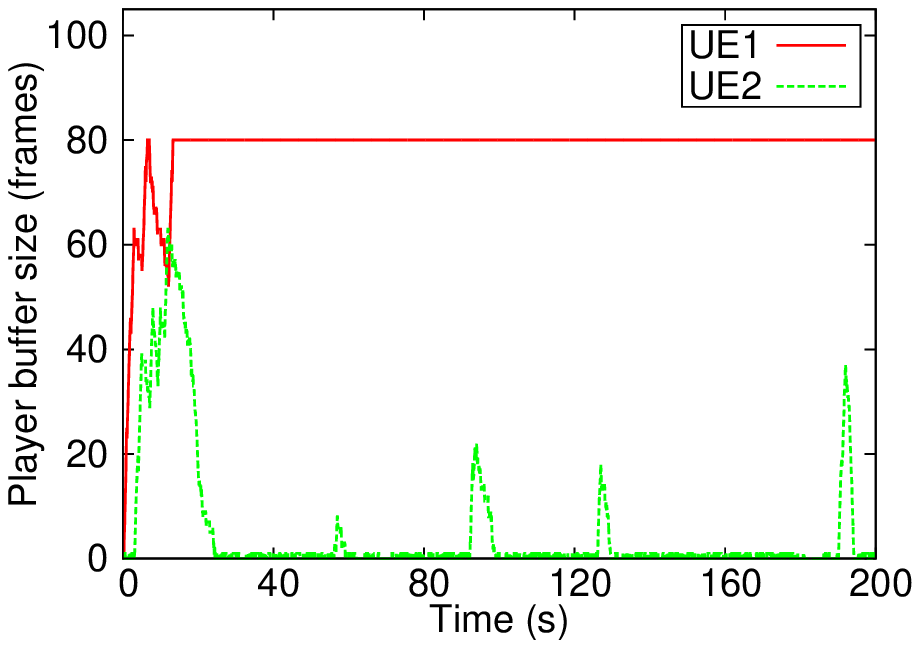}%
\label{buff_p}}
\hfil
\subfloat[]{\includegraphics[width=1.75in]{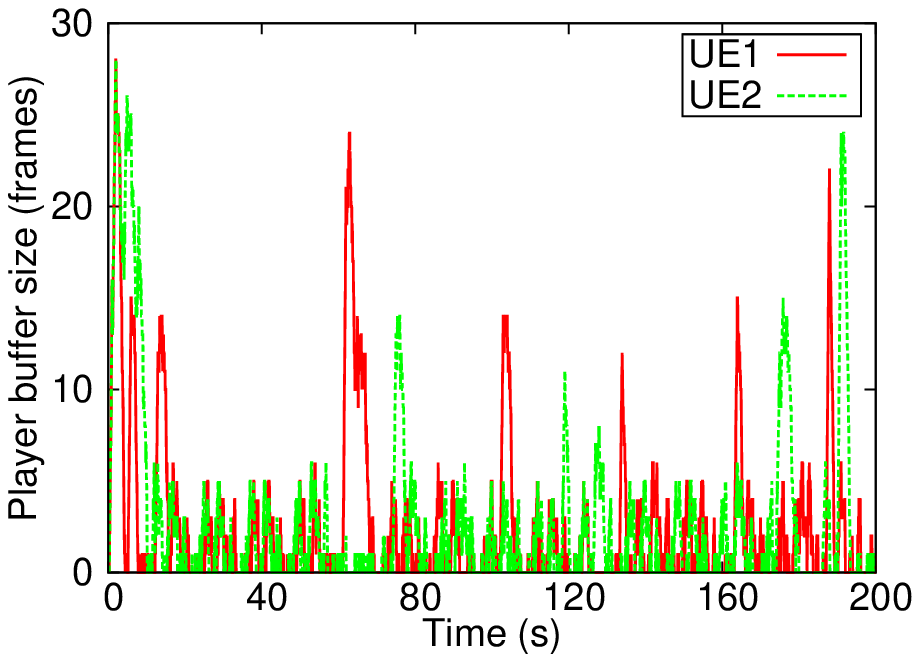}%
\label{buff_c}}}
\vspace{-2.5ex}
\caption{Buffer occupancy versus time of a) the proposed system; b) conventional client-centric system.}
\label{buff}
\vspace{-3.75ex}
\end{figure}

For buffer occupancy as shown in Fig. \ref{buff}, we observe that the high priority user never suffers buffer underflow in the network-assisted system even though congestion occurs. This is at the expense of performance degradation of the low priority user, which is acceptable based on service differentiation agreement. In the client-centric system, nonetheless, the competing players is frequently re-buffered.

\begin{table}[!t]\scriptsize
\renewcommand{\arraystretch}{1.3}
\caption{Playback Quality Metrics}
\vspace{-0.5ex}
\label{PQ}
\centering
\begin{tabular}{|c||c|c|c|}
\hline
UE1/UE2 & Average bit-rate  & Buffering ratio & Occurrence of buffering\\
\hline
Proposed & 303.95 / 150.18 (Kbps) & 0 / 17.71\% & 0 / 4.43 (per second)\\
\hline
Client-centric & 260.64 / 266.06 (Kbps) & 18.69\% / 18.79\% & 4.67 / 4.70 (per second)\\
\hline
\end{tabular}
\vspace{-4.5ex}
\end{table}

We also show industry-standard performance metrics \cite{QoESigcom13} in Table \ref{PQ}. We observe that the proposed system generally outperforms the client-centric system. It can also satisfactorily meet the service differentiation requirement.

\vspace{-1.75ex}
\section{Conclusion}\label{sec6}
\vspace{-1.5ex}
We proposed a generalized MDP-based adaptation framework for network-assisted mobile streaming by considering the playback, bandwidth, and economic factors in the mobile operator's view. With two case studies, the proposed framework is shown to outperform conventional client-side adaptation in terms of service provisioning. It also achieves near-ideal profit for the operator. Despite of potential complexity of MDP, the burden can be significantly released when implemented in the operator's proxy clouds. Future work can be focused on extending the framework to a larger user scale.

\bibliographystyle{IEEEtran}
\bibliography{IEEEabrv,F:/MyBib/wban/general,F:/MyBib/wban/PHY,F:/MyBib/mm_tx/DASH,F:/MyBib/mm_tx/qoe}


\end{document}